    \documentclass[a4paper,oneside,garamond]{article}
    \usepackage{amsmath}
    \usepackage{amssymb}
    \usepackage{latexsym}
    \usepackage{cite}
    \usepackage[dvips]{color}
    \usepackage{rotating}
    \usepackage[table]{xcolor}
\catcode`@=11

\@addtoreset{equation}{section}
\catcode`@=12

\title{
\textrm{Probing Psuedoscalars with Pulsar Polarisation Data Sets}}

\author{\textsf{\textbf{ Karam Chand$^a$, Subhayan Mandal$^a$}}\\
\normalsize
\textsl{$^a$Physics Department, Malaviya National Institute Of Technology, Jaipur, Rajasthan-302017,
India}\\
\normalsize
\texttt{e-mail addresses:
2015rpy9054@mnit.ac.in,
smandal.phy@mnit.ac.in}\\
}

\date{\today}

\begin{document}

\maketitle

\begin{center}
\begin{abstract}
\noindent
Recently a data set containing linear and circular polarisation information of a collection of six hundred pulsars has been released. 
The operative radio wavelength for the same was 21 cm. Pulsars radio emission process is modelled either with synchroton / superconducting self-Compton route 
or with curvature radiation route. These theories fall short of accounting for the circular polarisation observed, as they are predisposed towards producing, solely, linear polarisation. Here we invoke (pseudo)scalars 
and their interaction with photons mediated by colossal magnetic fields of pulsars, to account for the circular part of polarisation data. This enables us to estimate the pseudoscalar parameters such as its coupling to photons and its mass in conjunction as product. To obtain these values separately, we turn our attention to recent observation on 47 pulsars, whose absolute polarisation position angles have been made available. Except, a third of the latter set, the rest of it overlaps with the expansive former data set on polarisation type \& degree. This helps us figure out, both the pseudoscalar parameters individually, that we report here.  
\end{abstract}
\end{center}
\noindent
\section[int]{Introduction}
\begin{figure}[!h]
    \label{fig:ax-ph-mix}
    \includegraphics{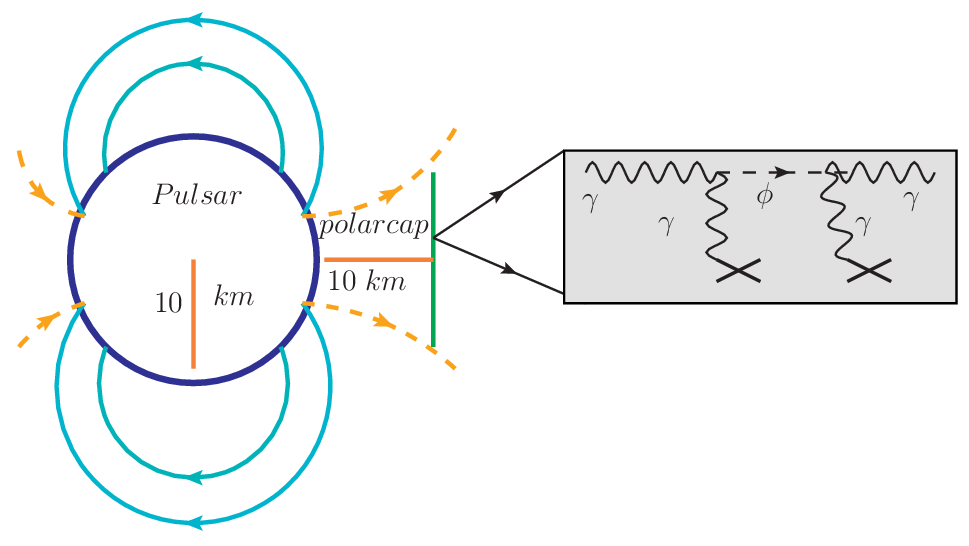}
    \caption[Polar Cap Conversion]{Axion Photon Mixing In The Polar Cap Region}
\end{figure}

In the last two decades, scenarios in which pseudoscalar \cite{Maiani86,pec77,wilc78,wein78,kim79,sik83} particles and photons couple and subsequently mix (fig no. \ref{fig:ax-ph-mix}) in the 
presence of magnetic fields have received a lot of attention \cite{Das05,saral04,agarwal11,agarwal12,jain13,tiwari12,tiwari16}, both phenomenologically \cite{Das08,payezprd12,payez11,vogel17,pelgrims15,pelgrims16,payez13,payez14} and observationally \cite{lamy00,hut98,hut01,hut05,hut14,hut10,jackson07,jagan16}. This 
is of particular interest in astrophysics, where this mixing of photons with pseudoscalars could make the universe transparent \cite{csaki03}, change 
the polarisation properties of light \cite{cudell08} and is be potentially responsible for effects such as ‘Supernovae dimming’ \cite{csaki03} or 
‘Large-scale coherent orientation’ \cite{cudell08} of the universe, also known as `Hutsemekers' effect. The best-known light pseudoscalar particle, 
the axion, was introduced long ago \cite{quinn77} to explain the absence of CP violation in Quantum Chromodynamics (QCD) \cite{pich95}. One 
postulated the existence of a new spontaneously broken continuous Peccei-Quinn symmetry, so that the axion was a pseudo Goldstone boson. It was soon 
realised that one needed to introduce a very large scale in the theory in order to suppress the interactions of the axion, while preserving the 
Peccei- Quinn mechanism. The invisible axion \cite{dine00} emerges at a unification scale, and the effective coupling is suppressed by this scale. 
The invisible axion, being closely related to QCD, has definite and interrelated expressions for its mass \cite{sikivie85} and coupling 
strength \cite{turner94} to other particles, given a specific model \cite{srednicki81,zakharov80}. Various cosmological and astrophysical bounds can 
be used to further constrain the parameters \cite{turner94}, and the allowed parameters do not lead to observable effects over cosmological scales. 
The mass of the pseudoscalar particle needs to be very close to the photon effective mass in order to mix in the rather weak magnetic 
fields of the extra galactic space. However, generic pseudoscalars or axion-like particles (ALP’s) have been hypothesized by many extensions of the 
standard model of particle physics. Theories such as Supergravity \cite{sengupta99} and Superstring theory \cite{sen01} contain many broken U(1) 
symmetries, that can lead to very light scalar, or pseudoscalar, particles.
\vskip 0.5cm
\par \noindent Pulsars, discovered fifty years back \cite{bell_68}, are a fusion fuel less state of a two to three solar mass 
$\left(M_{\odot}\right)$ star \cite{longair_94}, wherein surmounting inward gravitational pull \cite{longair_96}, in absence of a commensurate 
radiation pressure from fusion, makes it collapse \cite{pulsar}, into a tiny object \cite{lohfink_08}. Two effects follow: the protons and neutrons 
coalesce together making the Pulsars synonymous with neutron stars \cite{pul_hbk_05}; and, also during this compression phase the the magnetic flux is 
conserved, thereby promoting the magnetic induction field inside it to a colossal \cite{pul_hbk_05} value. Other effects such as the `pulsating' 
nature of the `star' in its last phase of stellar evolution, leading to the nomenclature \& discovery of the same \cite{pul_hbk_05}, won't be pursued 
here. 

Pulsars have been harnessed to estimate the coloumn density \cite{20_mil_pol_11}, by observing Pulsar dispersion measure. Also, the magnetic 
field of the interstellar medium (henceforth ISM) \cite{20_mil_pol_11} along the line of sight can be estimated by observing its rotation measure 
\cite{20_mil_pol_11}. Pulsars were traditionally, observed on earth inside the radio frequency window specifically, from 
100$\mathsf{MHz}\;-\;100\mathsf{GHz}$ \cite{pul_hbk_05}. However, over time, pulsars became known for emission in other wavebands like X-ray \& 
$\gamma$-ray, etc \cite{multi_pol_pul_17}. Despite, half a century of efforts, the mechanisms for such types of radiation and properties thereof, such 
as polarisation etc. are not very well understood \cite{pul_mag_I_17}. This in turn banks heavily on the fact that pulsar atmosphere or its 
magnetosphere models are still in its infancy \cite{pul_mag_II_16}. There are competing contenders as preferred models for pulsed emission and 
continuum radiation. Curvature radiation, synchroton radiation, inverse Compton radiation, superconducting self Compton radiation etc. are at the 
forefront, but none fits all observational features of pulsars \cite{pul_mag_I_17}. We shall, however, restrict ourselves, polarisation properties of 
radiation inside the pulsar atmosphere without looking into the radiation origin. Here we shall harness the two pulsar properties  the size, and magnetic field which in turn is deduced from period and associated derivative, to estimate the pseudoscalar parameters like the mass and its coupling to photons, with the help of 21 cm observations.
\vskip 0.5cm
\par \noindent In section no. \ref{data} we describe the polarimetric data set on six hundred pulsars \cite{parkesix18}, along with the quantities that can be 
derived from these observed parameters assuming a basic pulsar model \cite{pul_hbk_05}. The observation of circular polarisation is hitherto 
unexplained by radiation models, theoretical \cite{wang12,ganga10,ardavan08} \& statistical \cite{mckinn02} alike, so far. Thereafter, in the section no. \ref{ppm} , we invoke the light quanta to pseudoscalar 
interaction to step wise calculate the correlators, ab initio, between the three degrees of freedom. Thereby, in section no. \ref{sp} we digress to 
stokes parameters; the experimental interface with theoretical quantities like ellipticity parameter and polarisation position angle, using the definition of correlators. In 
the next section \ref{epppa}, we discuss the extraction process of pseudoscalar parameters, for mixing case only, discarding another two cases and leaving the general case open that might arise, naturally. Also in this segment we estimate the values of 
regression parameters derived from statistical analysis of the data set tables. In the next section we present our results. Thereafter, we conclude by projecting the feasibility of our result and scope in future directions. 
\section[data]{Observation}
\label{data}
The following data shown in table no. (\ref{bigtable}) is a small part of the data obtained from the reference no. \cite{parkesix18}. It 
contains the spin down luminosity $\dot{E}$ \cite{pulsar} and pulsar spin period ${P}$ \cite{lohfink_08} all six hundred of them. Following the 
basic pulsar model \cite{diagtool13,pul_hbk_05} we may derive pulsar parameters such as spin period time derivative $\dot{P}$ and the magnetic field 
$B_{s}$.
\begin{eqnarray}
    \label{eq:magfeld}
  B_s &=& 3.2\times10^{19}\times\sqrt{P\dot{P}}\nonumber \\
  \dot{P} &=& \frac{\dot{E}P^3}{4\pi^2I}
\end{eqnarray}
 
\par\noindent Also, from the ratio between the percentage of circular to linear polarisation provides us with the ellipticity parameter $\chi$. 
\begin{equation}
  \tan\left(2\chi\right) = {V\over p_{lin}}
\end{equation}
We have extracted \& separately tabulated these derived values for further use in section no. \ref{epppa}.

\begin{sidewaystable}
\caption{Pulsar polarisation properties at 1.4~GHz (sample)}
\label{bigtable}
\begin{tabular}{lrrcccccrr}
Jname & Period & $\dot{P}$ & log($\dot{E}$) & $L/I$ & $V/I$ & $|V|/I$ & err & B & $\chi$ \\
& (ms) & unitless & (ergs$^{-1}$)  & \% & \% & \% & \% & Gev$^2$ & rad \\
\hline & \vspace{-3mm} \\
J0034-0721&943&4.24E-16&31.3&10.7&7.7&7.5&3&4.44E-08&0.303493832023204 \\
J0051+0423&354.7&7.14E-18&30.8&13.1&-2.3&11.2&3.3&3.53E-09&0.287272232961884 \\
J0108-1431&807.6&8.43E-17&30.8&76.7&15.5&13.1&3.1&1.83E-08&0.085016451375219 \\
J0134-2937&137&8.21E-17&33.1&45.3&-17.2&16.9&3&7.44E-09&0.178538083464964 \\
J0151-0635&1464.7&3.99E-16&30.7&29.1&-1.7&4.2&3.3&5.36E-08&0.070948527302082 \\ 
J0152-1637&832.7&1.16E-15&31.9&15.1&1.1&6&3&6.90E-08&0.190253188556182 \\
J0206-4028&630.6&1.27E-15&32.3&10.6&9.3&9.9&3.1&6.27E-08&0.366407550893253 \\
J0211-8159&1077.3&3.17E-16&31&17&11.7&15.4&5.5&4.10E-08&0.502033554635695 \\
J0255-5304&447.7&2.86E-17&31.1&7.3&-4.1&5.5&3&7.94E-09&0.329545023167104 \\
J0304+1932&1387.6&1.35E-15&31.3&33.4&15.1&14.8&3&9.60E-08&0.209112164789615 \\
J0343-3000&2597&5.59E-17&29.1&14.3&3.1&3.9&3.2&2.67E-08&0.107846382092558 \\
J0401-7608&545.3&1.64E-15&32.6&28.6&-0.1&4.7&3&6.63E-08&0.08060916072996 \\
J0448-2749&450.4&1.46E-16&31.8&23.9&-13.3&11.8&3&1.80E-08&0.225223527837328 \\
J0450-1248&438&1.07E-16&31.7&25.3&2.5&6&3.4&1.52E-08&0.126047990334037 \\
J0452-1759&548.9&5.28E-15&33.1&18.9&3.6&4.2&3&1.19E-07&0.109334472936971 \\
J0459-0210&1133.1&1.47E-15&31.6&10.4&-12.9&9.6&3.8&9.05E-08&0.337370471111776 \\
J0520-2553&241.6&2.84E-17&31.9&18.2&-4.3&5&3.6&5.81E-09&0.129970070094264 \\
J0525+1115&354.4&7.12E-17&31.8&10.6&12.5&15.5&3&1.11E-08&0.478954335502063 \\
J0528+2200&3745.5&4.21E-14&31.5&36.9&-4.9&4.6&3&8.81E-07&0.062683228726001 \\
J0533+0402&963&1.8E-16&30.9&13.3&4&5.5&3.2&2.92E-08&0.211767678516289 \\
J0536-7543&1245.9&6.17E-16&31.1&48.8&-11.1&11&3&6.15E-08&0.110852247980192 \\
J0540-7125&1286&8.55E-16&31.2&14.2&3.1&15.8&4.8&7.35E-08&0.391121799864486 \\
J0543+2329&246&1.5E-14&34.6&45.2&-8.2&7.9&3&1.35E-07&0.086515496781741 \\
J0601-0527&396&1.25E-15&32.9&30.9&4.4&11.3&3&4.94E-08&0.175294355633881 \\
J0614+2229&335&6.01E-14&34.8&72&20.3&20.1&3&3.15E-07&0.135938275210493 \\

\hline
\end{tabular}
\end{sidewaystable}
\section{Pseudoscalar Photon Mixing}
\label{ppm}
The following mixing matrix (eqn. no. \ref{eq:matrix00}) provides for the necessary ingredient of photon pseudoscalar mixing mediated by a magnetic 
field \cite{stodol88}. Also this reference assumes a free space, for calculation, hence there are no Faraday effect ($M_{21}$ \& $M_{12}$ entries) 
terms coupling the two photon polarisations. Inside pulsar magnetosphere this could hardly be the case. However, we may still ignore the Faraday 
terms. 
The reason being the smallness of it inside spaces with large magnetic fields; as shown in by one of the coauthors \cite{mandal09}, by deriving the 
limiting propagation frequency, below which Faraday effect holds significance.
\begin{displaymath}
  \omega_L = \frac{\omega_p^2\omega_B\left(\omega_p^2-m_a^2\right)}{|\mathcal{B}|^2}{\cos\Theta\over\sin^2\Theta}
\end{displaymath}
for the values derived from the pulsar database, such as the magnetic field, and the plasma frequency from literature \cite{melrose17}, which is much 
smaller than the pseudoscalar mass, we see that Faraday effect can safely be neglected at the operating frequency of 1.4 GHz ($\gg \omega_L$), with 
which 
the observations were made.
\renewcommand{\thefootnote}{(\arabic{footnote})}
\begin{equation}
M = \left(\begin{matrix}A_1 & 0 & 0\cr
           0 & A_2 & T\cr
	   0 & T  & -B\end{matrix}\right)\,,
\label{eq:matrix00}
\end{equation}

\noindent
Where the symbols in the matrix [{\ref{eq:matrix00}}] stands for 
\begin{eqnarray}
 A_1&=&4\omega^2\xi Sin^2 \Theta +\omega_p^2 \nonumber  \\
 A_2&=&7\omega^2\xi Sin^2 \Theta +\omega_p^2, \nonumber\\ 
B&=&m_a^2 , \\ 
T&=& g \mathcal{B}\omega.\nonumber \\ 
\label{symb1} \nonumber
\end{eqnarray}
\noindent
where, $\mathcal{B}$ is the magnetic field, $\Theta$ is the angle between the  $\vec{k}$ and the magnetic field $\mathcal{B}$, $m_a$ the axion mass, and $\xi=
{\alpha\over45\pi}\left(\frac{e\mathcal{B}}{m^2_f}\right)^2$, with $m_f$ \cite{stodol88} the lightest Fermion mass. 

\noindent
The non diagonal 2{\bf{x}}2 matrix, in eqn.{\ref{eq:matrix00}} is given by,

\begin{equation}
M_2 = \left(\begin{matrix}A_2 & T \cr
                           T &  -m_a^2 
	   \end{matrix}\right)\,.
\label{eq:matrix10}
\end{equation}

One can solve for the eigen values of the eqn. [\ref{eq:matrix10}], from the determinant equation,

\begin{equation}
 \left|\begin{matrix}A_2-\lambda & T \cr
                           T &  -m_a^2-\lambda 
	   \end{matrix}\right| = 0, 
\label{eq:matrix10}
\end{equation}
\noindent
and the roots are,\\

\begin{eqnarray}
M_{\pm}= \frac{A_2-m_a^2}{2}
\pm \frac{1}{2}\sqrt{\left[(A_2+m_a^2)^2 + 4 T^2\right]}.\\
\end{eqnarray}

\subsection{Equation of Motion}
\label{eom}

The equation of motion for the axion photon mixing, in the non diagonal
basis gets decoupled and can be written
in the matrix from as:,
\begin{equation}
\left[(\omega^2 + \partial_z^2) \,\,{\bf I} +\mbox{ \bf{M} }  \right]
\left( \begin{array}{c}
A_{\perp}  \\
A_{\parallel}\\
a      \\
\end{array} \right)=0.
\end{equation}
where ${\bf I}$ is a $3\times 3$ identity matrix and {\bf{M}} is the mixing
 matrix.\\

\noindent
The uncoupled and the coupled  equations can further be written as,
\begin{equation}
\left[(\omega^2 + \partial_z^2)  +\mbox{A}_{1}  \right]
\left( \begin{array}{c}
A_{\perp}  \\
\end{array} \right)=0.
\end{equation}
and
\begin{equation}
\left[(\omega^2 + \partial_z^2) \,\,{\bf I} +  M_2  \right]
\left( \begin{array}{c}
A_{\parallel}\\
a      \\
\end{array} \right)=0.
\label{c:eqn}
\end{equation}\\

\noindent
It is possible to diagonalise eqn.[\ref{c:eqn}] by a similarity 
transformation (we would denote the diagonalising matrix by $O$),  
leading to the form,\\

\begin{equation}
\left[(\omega^2 + \partial_z^2) \,\,{\bf I} +\mbox{ \bf{$M_D$} }  \right]
\left( \begin{array}{c}
\bar{A}_{\parallel}\\
\bar{a}      \\
\end{array} \right)=0.
\label{c:eqnd}
\end{equation}
when the diagonal matrix $M_D$ is given by:

\begin{eqnarray}
 M_D= 
   \left(\begin{matrix}M_{+} & 0 \cr
                           0 & M_{-} 
	   \end{matrix}\right)
\label{eq:matrix15}
\end{eqnarray}\\

\subsection{Dispersion Relations}
\label{dr}

\noindent
Defining the wave vectors in terms of $k_{i}$'s, as: 
\begin{eqnarray}
k_{\perp}&=& \sqrt{\omega^2+ A_1 } \nonumber \\
k_{+}&=&\sqrt{\omega^2+M_{+}} \nonumber \\
k_{-}&=&-\sqrt{\omega^2+M_{+}}
\label{pos}
\end{eqnarray}

and
\begin{eqnarray}
k'_{+}&=&\,\,\sqrt{\omega^2+M_{-}} \nonumber \\
k'_{-}&=&-\sqrt{\omega^2+M_{-}}
\label{neg}
\end{eqnarray}

\subsection{Solutions}
\label{sol}

\noindent
The solutions for the gauge field and the axion field, given by [\ref{c:eqnd}]
as well as the solution for eqn. for $A_{\perp}$ in $k$ space can be written as,
\begin{eqnarray}
\bar{A}_{||}(z)&=&\!\! \bar{A}_{||_{+}}(0)e^{ik_{+}z}+\bar{A}_{||_{-}}(0)e^{-ik_{-}z}\\
\bar{a}(z)&=&\bar{a}_{+}(0)\,\,e^{ik'_{+}z}+\bar{a}_{-}(0)\,\,e^{-ik'_{-}z}\\
A_{\perp}(z)&=&\!\! A_{{\perp}_{+}}(0)e^{ik_{\perp}z}+A_{{\perp}_{-}}(0)
e^{-ik_{\perp}z}\\
\end{eqnarray}

\noindent
The diagonal matrix can be written as
\begin{eqnarray}
M_{D}=O^{T} M_2 O
\end{eqnarray}
when
\begin{equation}
 O=\left(\begin{matrix}\cos\theta & -\sin\theta \cr
                           \sin\theta &  \cos\theta  
	   \end{matrix}\right) \equiv 
\left(\begin{matrix}c & -s \cr
                           s &  c  
	   \end{matrix}\right). 
\label{eq:matrix14}
\end{equation}
in short hand notation.\\

\subsection{Similarity Transformation}
\label{st}
\noindent
The diagonal matrix
\begin{equation}
 M_D= 
\left(\begin{matrix}c & s \cr
                           -s &  c  
	   \end{matrix}\right)
   \left(\begin{matrix}M_{11} & M_{12} \cr
                           M_{21} & M_{22} 
	   \end{matrix}\right)\left(\begin{matrix}c & -s \cr
                           s &  c  
	   \end{matrix}\right), 
\label{eq:matrix15}
\end{equation}\\

\noindent
With $M_{11}=A_2 $, $M_{12}=T$, $M_{21}=T$ lastly $M_{22}=-B$.\\

\noindent
The value of the parameter $\theta$, is fixed from the equality,
\begin{equation}
 M_D= 
\left(\begin{matrix}c & s \cr
                           -s &  c  
	   \end{matrix}\right)
   \left(\begin{matrix}M_{11} & M_{12} \cr
                           M_{21} & M_{22} 
	   \end{matrix}\right)\left(\begin{matrix}c & -s \cr
                           s &  c  
	   \end{matrix}\right)= \left(    
                                 \begin{matrix} M_{+} & 0 \cr
                                                0     & M_{-} \cr
                                 \end{matrix} 
                                 \right), 
\label{eq:matrix17}
\end{equation}\\
\noindent
leading to,

\begin{equation}
\left(
      \begin{matrix}
        c^2 M_{11}+s^2 M_{22}+ 2csM_{12} & M_{12}(c^2-s^2)+cs(M_{22}-M_{11}) \cr
        M_{12}(c^2-s^2)+cs(M_{22}-M_{11}) &  s^2 M_{11}+c^2 M_{22}- 2csM_{12}  
      \end{matrix}
\right)=                        \left(    
                                 \begin{matrix} M_{+} & 0 \cr
                                                0     & M_{-} \cr
                                 \end{matrix} 
                                \right), 
\label{eq:matrix17}
\end{equation}\\
\noindent
Equating the components of the matrix equation [\ref{eq:matrix17}], one arrives at:
\begin{eqnarray}
\rm{tan}(2\theta)= \frac{2M_{12}}{M_{11}-M_{22}}=\frac{2T}{A_2-m^2_a}.
\label{eq:tantheta}
\end{eqnarray}\\

\subsection{Correlation Functions}
\label{cf}

\noindent
The solutions for propagation along the $+$ve z axis, is given by,
\begin{eqnarray}
\bar{A}_{||}(z)&=&\!\! \bar{A}_{||}(0)e^{ik_{+}z}\\
\bar{a}(z)&=&\bar{a}(0)\,\,e^{ik'_{+}z}\\
\end{eqnarray}
\noindent
that can further be written in the following form,
\begin{eqnarray}
\left(
      \begin{matrix}
                 \bar{A}_{||}(z) \cr
                 \bar{a}(z)
      \end{matrix}
\right)
=
\left(
  \begin{matrix}
       e^{ik_{+}z} & 0 \cr
       0           & e^{ik'_{+}z}
  \end{matrix}
\right)
\left(
   \begin{matrix}
     \bar{A}_{||}(0) \cr
     \bar{a}(0)
   \end{matrix}
\right)
\label{eq:matrix21}.
\end{eqnarray}
\noindent
Since,
\begin{eqnarray}
\left(
   \begin{matrix}
     \bar{A}_{||}(z/0) \cr
     \bar{a}(z/0)
   \end{matrix}
\right)
=
O^{T}\left(
   \begin{matrix}
     {A}_{||}(z/0) \cr
     {a}(z/0)
   \end{matrix}
\right)
\label{eq:matrix23}.
\end{eqnarray}\\

\noindent
it follows from there that,\\

\begin{eqnarray}
\left(
   \begin{matrix}
     A_{||}(z) \cr
     a(z)
   \end{matrix}
\right)
=
O
\left(
\begin{matrix}
e^{ik_+z} & 0            \cr
0         & e^{ik'_{+}z}
\end{matrix}
\right)
O^T
\left(
   \begin{matrix}
     {A}_{||}(0) \cr
     {a}(0)
   \end{matrix}
\right)
\label{eq:matrix23}.
\end{eqnarray}

Using eqn.[\ref{eq:matrix23}] we arrive at the relation,
\begin{eqnarray}
A_{||}(z)= \left[ e^{ik_{+}z} \rm{cos}^2\theta + e^{ik'_{+}z} \rm{sin}^2\theta
            \right]A_{||}(0) +                
\left[ e^{ik_{+}z}- e^{ik'_{+}z}\right] 
\rm{cos}\theta \,\rm{sin}\theta \,a(0) \\
 a(z)= \left[ e^{ik_{+}z}- e^{ik'_{+}z}\right] \rm{cos}\theta \, \rm{sin} \theta \, 
A_{||}(0) + 
\left[ e^{ik_{+}z} \rm{sin}^2\theta + e^{ik'_{+}z} \rm{cos}^2\theta \right] a(0) 
\end{eqnarray}

If the axion field is zero to begin with, i.e 
\begin{equation}
a(0)=0.
\end{equation}

\noindent
Then the solution for the gauge fields take the following form,
\begin{eqnarray}
A_{||}(z)\!\!\!\!&=&\!\!\!\!\! \left[ e^{ik_{+}z} \rm{cos}^2\theta + e^{ik'_{+}z} \rm{sin}^2\theta
            \right]A_{||}(0) \\
A_{\perp}(z)\!\!\!\!\!&=&\!\!\!\!\!e^{ik_{\perp}z}A_{\perp}(0).
\end{eqnarray}

\noindent
The correlations of different components take the following form:
\begin{eqnarray}
<A^{*}_{||}(z)A_{||}(z)>\!\! &=& \!\!\left[ \rm{cos}^{4}\theta +\rm{sin}^4\theta + 2\, \rm{sin}^2\theta \,\,
\rm{cos}^2{\theta}\,\, \rm{cos}\left[\left(k_{+}-k'_{+}\right)z\right]\, \right]<A^{*}_{||}(0)A_{||}(0)> \\
<A^{*}_{||}(z)A_{\perp}(z)>\!\!&=&\!\! \left[ \rm{cos}^2\, \theta e^{i(k_{\perp}-k_{+})z}+ \rm{sin}^2 \theta \,
e^{i(k_{\perp}-k'_{+})z} \right]\,<A^{*}_{||}(0)A_{\perp}(0)>\\
<A^{*}_{\perp}(z)A_{\perp}(z)>\!\!&=&\!\!<A^{*}_{\perp}(0)A_{\perp}(0)>
\end{eqnarray}

\section{Stokes Parameters}
\label{sp}
Using the definitions of the Stokes parameters, in terms of the correlators:
\begin{eqnarray}
\rm{I}\!\!&=&\!\!<A^{*}_{||}(z)A_{||}(z)>+<A^{*}_{\perp}(z)A_{\perp}(z)>, \\
\rm{Q}\!\!&=&\!\!<A^{*}_{||}(z)A_{||}(z)>- <A^{*}_{\perp}(z)A_{\perp}(z)>, \\
\rm{U}\!\!&=&\!\!2 \rm{Re}<A^*_{||}(z)A_{\perp}(z)>, \\
\rm{V}\!\!&=&\!\!2\,\rm{Im}<A^*_{||}(z)A_{\perp}(z)>.
\end{eqnarray}\\

\noindent
Using the relations for the corresponding correlators, the stokes parameters turn out to be

\begin{eqnarray}
\rm{I}\!\!&=&\!\! \left[ \rm{cos}^{4}\theta +\rm{sin}^4\theta + 2\, \rm{sin}^2\theta \,\,
\rm{cos}^2{\theta}\,\, \rm{cos}\left[\left(k_{+}-k'_{+}\right)z\right]\, \right]<A^{*}_{||}(0)A_{||}(0)>+
<A^*_{\perp}(0)A_{\perp}(0) >  \nonumber \\
\rm{Q}\!\!&=&\!\!\left[ \rm{cos}^{4}\theta +\rm{sin}^4\theta + 2\, \rm{sin}^2\theta \,\,
\rm{cos}^2{\theta}\,\, \rm{cos}\left[\left(k_{+}-k'_{+}\right)z\right]\, \right]<A^{*}_{||}(0)A_{||}(0)>-
<A^*_{\perp}(0)A_{\perp}(0)>  \nonumber \\
\rm{U}\!\!&=&\!\! 2\left( \left[ \rm{cos}^2\theta \, \rm{cos}\left[ \left(k_{\perp} - k_{+}\right) z\right]\right]
+ \rm{sin}^2 \theta \rm{cos}\left[\left(k_{\perp}- k'_{+}\right)z \right] \right) <A^{*}_{||}(0)A_{\perp}(0)> \nonumber \\
\rm{V}\!\!&=&\!\! 2\left( \left[ \rm{cos}^2\theta \, \rm{sin}\left[\left(k_{\perp} - k_{+}\right) z\right] \right]
+ \rm{sin}^2 \theta \rm{sin}\left[\left(k_{\perp}- k'_{+}\right)z \right] \right) <A^{*}_{||}(0)A_{\perp}(0)>
\label{stks}
\end{eqnarray}

\noindent
The Stokes parameters are also expressed as such
\begin{eqnarray}
\rm{I}&=&\rm{I}_p \\
\rm{Q}&=&\rm{I}_p \rm{cos}2\psi \rm{cos}2\chi \\
\rm{U}&=&\rm{I}_p \rm{sin}2\psi \rm{cos}2\chi \\
\rm{V}&=&\rm{I}_p  \rm{sin}2\chi.
\label{stks}
\end{eqnarray}
where $\chi$ \& $\psi$ are are usual ellipticity parameter and the polarisation position angle.
\noindent
The degree of (linear /) polarisation is given by,
\begin{eqnarray}
p&=&\frac{\sqrt{\rm{Q}^2+\rm{U}^2+\rm{V}^2}}{\rm{I}_P} \nonumber \\
p_{lin} &=& \frac{\sqrt{\rm{Q}^2+\rm{U}^2}}{\rm{I}_P}
\end{eqnarray}
and the linear polarisation angle is given by
\begin{eqnarray}
    \label{eq:defn}
\rm{tan}2 \psi &=& \frac{U}{Q}\nonumber \\
\rm{tan}2 \chi &=& \frac{V}{p_{lin}}
\end{eqnarray}
\par\noindent It has been noted in \cite{ganguly12}, that in case, we make any coordinate transformation around the axis of photon propagation the 
two linear polarisation become mixed. Hence, we need to be careful, as our solution process entails a similarity transformation. To see this we 
define the density matrix
\begin{equation}
 \rho(z) = \left(
  \begin{matrix}
       \left<A^{*}_{||}(z)A_{||}(z)\right> & \left<A_{||}(z)A_{\perp}^{*}(z)\right> \cr
       \left<A^{*}_{||}(z)A_{\perp}(z)\right>         & \left<A^*_{\perp}(z)A_{\perp}(z)\right>
  \end{matrix}
  \right) = {1\over2}\left(
  \begin{matrix}
       I(z)+Q(z) & U(z)-iV(z) \cr
       U(z)+iV(z)  & I(z)-Q(z)
  \end{matrix}
\right)
\end{equation}
\par\noindent if we rotate the density matrix by an amount $\alpha$ about an axis perpendicular the plane containing $A_{||}(z)$ \& $A_{\perp}(z)$, 
the density matrix transforms as $\rho(z) \longrightarrow \rho^{'}(z)$ given such to be
\begin{equation}
  \rho{'}(z) = {1\over2}R(\alpha)\left(
  \begin{matrix}
       I(z)+Q(z) & U(z)-iV(z) \cr
       U(z)+iV(z)  & I(z)-Q(z)
  \end{matrix}
  \right)R^{-1}(\alpha)
\end{equation}
\par\noindent where,
\begin{equation}
 R(\alpha)= \left(\begin{matrix}\cos\alpha & -\sin\alpha \cr
                           \sin\alpha &  \cos\alpha  
	   \end{matrix}\right)
\end{equation}
\par\noindent Under such transformation the I(z) \& V(z) remains unaltered. However, the Q(z) \& U(z) starts mixing with each other by the following 
\begin{equation}
\begin{pmatrix}
  Q^{'}(z) \\
  U^{'}(z)
\end{pmatrix}
=\begin{pmatrix}
\cos 2\alpha & \sin2\alpha \\
-\sin 2\alpha & \cos2\alpha
\end{pmatrix}\begin{pmatrix}
  Q(z) \\
  U(z)
\end{pmatrix}
\end{equation}
We conclude this section by mentioning that in such a case the ellipticity parameter remains unaltered but the polarisation position angle changes by 
$2\alpha$ as given below
\begin{eqnarray}
    \label{roteff}
  \tan\left(2\chi{'}\right) &=& \tan\left(2\chi\right) \nonumber \\
  \tan\left(2\psi{'}\right) &=& \tan\left(2\alpha+2\psi\right)
\end{eqnarray}

\section{Ellipticity Parameter \& Polarization Position Angle}
\label{epppa}
\noindent
As a follow-up to the analytical expressions given in the previous section/s, we consider two special case of the stokes parameter where either one 
of the two effects, namely, the mixing effect or, the vacuum birefringence effect would be absent. Thereafter we shall consider the general formula. In each 
case, we would like to obtain the value of the ellipticity angle $\chi$ after propagation a fixed distance $z$ of light and determine it's frequency 
dependence. For all the three cases we shall assume the light to be completely plane polarised in the transverse direction, or $U$ polarised. This is 
common observance in Pulsar polarisation cases.
\subsection{Case - I: Mixing Only}
\par\noindent
Here we assume that the vacuum birefringence terms (i.e. $\xi$ term inside the diagonal ones $A_1,A_2$) are absent. We also assume a pseudoscalar mass which 
is much less than the plasma frequency here. This greatly simplifies calculation without being much deviant from the reality, if we consider the 
parameters of the pulsar environment. Next we consider how the circular polarisation varies in 
this case. Assuming $\theta \ll 1$ one have 
\par\noindent
\begin{equation}
\rm{V}\!\!=\!\! \left(\rm{sin}\left[\left(k_{\perp} - k_{+}\right) z\right] + \left[\frac{g\mathfrak{B}\omega}{  
\omega_p^2 + m_\mathfrak{a}^2}\right]^2\rm{sin}\left[\left(k_{\perp} - k_{+}^{'}\right) z\right]\right) <A^{*}_{||}(0)A_{\perp}(0)>
\label{circx}
\end{equation}
\par\noindent
Following the set of eqns. \ref{pos}-\ref{neg} we can simplify the arguments of the remaining sinusoids of eqn. no. \ref{circx} as given below :
\begin{equation}
  k_{\perp} - k_{+} = - \;\left\lbrace\,{\left(g\mathfrak{B}\omega\right)^2 \over 2\left(\omega_p^2 + m_\mathfrak{a}^2\right)\omega}\,\right\rbrace 
\;\;\;
k_{\perp} - k_{+}^{'} = +\;\left\lbrace\, {\left(g\mathfrak{B}\omega\right)^2 \over 2\left(\omega_p^2 +
m_\mathfrak{a}^2\right)\omega}+{m_\mathfrak{a}^2 \over 2\omega}\,\right\rbrace
\label{simx}
\end{equation}
\par\noindent
So, if $\xi = 0$, then the ellipticity parameter to its lowest order ($\propto \theta^2$ ) is found to be as follows, which matches well with 
\cite{cameron99,ganguly12}, though the later most probably has a typo \footnote{It claimed concurrence with the former but is actually at variance, with 
it}.
\begin{equation}
    \label{eq:ellp}
  \chi \approx {1\over96\omega}\left(g\mathfrak{B}m_a\right)^2z^3
\end{equation}
Similarly, we may now turn our attention to two linear polarisation degrees of freedom, where the mixing angle $\theta \ll 1$, is small, to figure out the polarisation position angle.
\begin{equation}
 \tan(2\psi)=\frac{U}{Q}
\end{equation}
However, in the beginning of this section we have already mentioned that $U \simeq 1$. This is true for the parameters of interest used here and the observational cases to be discussed later. This makes the polarisation position angle inversely proportional to $Q$. But before we evaluate the expression for $Q$, we note that in the case of mixing the beam is assumed to propagate at an angle $\pi\over4$ as compared to the magnetic field of the Pulsar. Hence we need to change our expression for polarisation position angle accordingly. As discussed during derivation of eqn. no. (\ref{roteff}), we have;
\begin{equation}
    \tan\left(2\psi+{\pi\over2}\right)=\frac{1}{Q}
\end{equation}
Next, we evaluate $Q$ keeping in mind the approximations made before. Keeping terms up to order $\theta^2$ in the expression for $Q$, we have,
\begin{equation}
    {Q}= -2\theta^2\left[\sin^2\left(\frac{\left\langle k_+ - k_+^{'}\right\rangle z}{2} \right)\right]
\end{equation}
Again, following the set of Eqns. \ref{pos}-\ref{neg}, we have 
\begin{equation}
 \left\langle k_+ - k_+^{'}\right\rangle \simeq \frac{m_a^2}{2\omega}   
\end{equation}
Substituting, one gets, in conjuntion with \cite{cameron99}
\begin{equation}
    \psi = {1\over16}\left(g\mathfrak{B}z\right)^2
\end{equation}
However, unlike the circular polarisation, which was attributed to its entirety, to the mixing effect, one  can not ascribe the entire pulsar linear polarisation \cite{gedal02} to this tiny mixing effect, where the mixing angle $\theta \ll 1$. So, we note that the pulsar radio emission is inherently linearly polarised to a large degree, due to curvature, synchroton and superconducting self Compton effects thereof. We use $U \simeq 1$ and only the $Q$ part is modelled via pseudoscalar photon mixing; where  
\begin{equation}
 Q=\frac{1}{8}\left(g\mathfrak{B}z\right)^2
\end{equation}
along with the definitive couple of eqn. (\ref{eq:defn}) to note that the linear polarisation observed is equal to 
\begin{equation}
    p_{lin}=Q\,\sec\left(2\psi+{\pi\over2}\right)
\end{equation}
We note that the determination process of absolute pulsar polarisation \cite[absolute PPAs]{rank15} position angles, is now experimentally feasible and the same values have already been scraped out for 30 (thirty) odd pulsars. The literature contains a little less than fifty absolute PPAs from \cite[absolute polarisation position angles for 47 odd pulsars]{ranki15}, out of which only 30 (thirty) cross matched with that of our old set of 537 data, used to calculate the ellipticity parameter. The expression for $Q$ has only one unknown, the coupling of pseudoscalar with photons. Hence, we may do a regression analysis here, too, to estimate the same. The summary table is given in table no. (\ref{regres}).
\begin{figure}[t]
    \label{fig:reg}
    \includegraphics{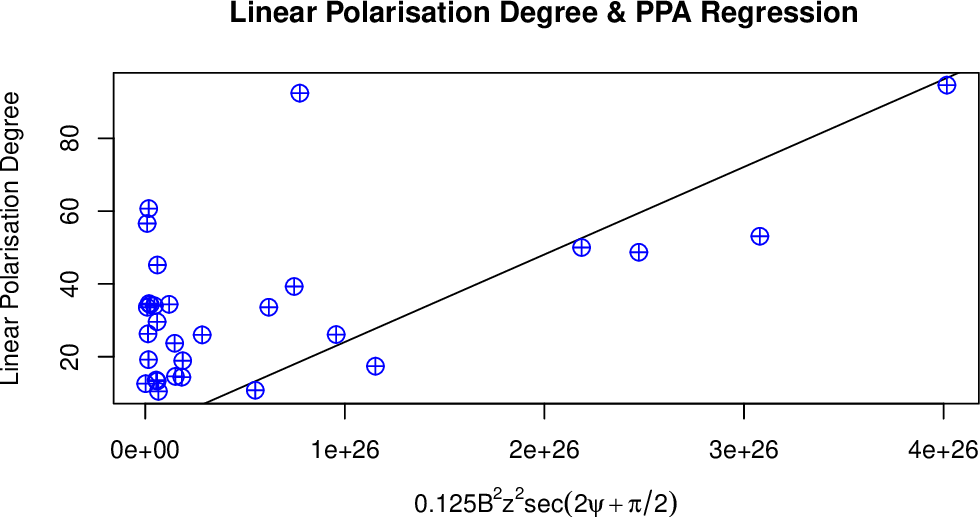}
    \caption[PPA vs. LinPol]{Linear regression between PPAs and Lin. Pol. Abscissa is in GeV$^2$ units \& the ordinate is dimensionless.}
\end{figure}
\begin{table}
\begin{center}
\begin{tabular}[!h]{lccccr}
Coefficients & Mean & Std.-Error & F-statistics & t-value & Pr$(>|t|)$ \\
 \hline
 Slope & 2.404e-25 & 4.567e-26 & 27.7196 & 5.265 & 1.21e-05\\
 \hline
 \end{tabular}
 \caption{Regression Result for the coupling of the pseudoscalar}
 \label{regres}
\end{center}
\end{table}

\begin{table}[h]
\begin{center}
\caption{Small Sample of Absolute Pulsar Polarisation Position Angles, from \cite{ranki15}.\label{tbl-1}}
\smallskip
\footnotesize
\begin{tabular}{llllll}
\hline
Pulsar & $PA_V$ & $PA_0$  & $\Psi$  \\
 & (deg) && (deg)  \\ 
\hline
 \\
B0011+47 & +136(3) & 43(7) & --87(8) \\ 
B0136+57 & --131(0) & 43(3) & 6(3) \\
B0329+54 & 119(1) &  20(4)  &   99(4)  \\
B0355+54 &  48(1) &  --41(4) &   89(5)  \\ 
\medskip
B0450+55 & 108(0) & --23(16)& --94(16) \\ 
B0450--18 & 40(5) & 47(3) & --7(6) \\
B0540+23 &  58(19) & --85(3) & --37(19) \\
B0628--28 & 294(2) &  26(2) & 88(3) \\
B0736--40 & 227(5) & --44(5) & 91(7) \\  
\hline
\end{tabular}
\end{center}
\end{table}
For the sake of brevity, we post a small segment of total 47 pulsar given in reference \cite{ranki15} in table no. (\ref{tbl-1}). The pulsar names here are catalogued in B1950 almanac standard, which were then converted to J2000 almanac standard and cross matched with the original \& usable 537 strong population data on pulsar polarisation. Thirty (30) odd samples of them were found to be common in both. 
\begin{figure}[!h]
    \label{fig:regg}
    \includegraphics[width=\textwidth]{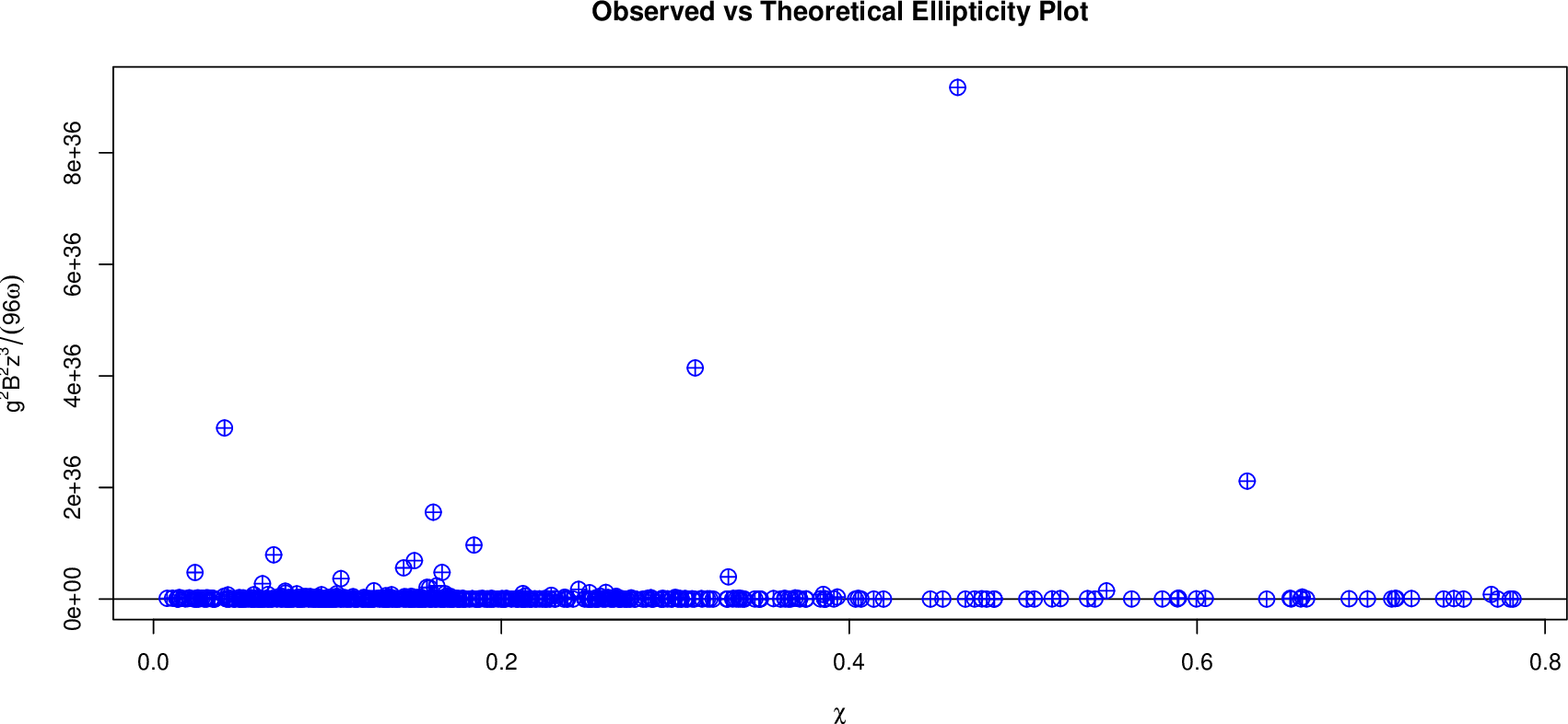}
    \caption[Ellipticity parameter vs. Circular Polarisation]{Linear regression between observed \& (scaled) theoretical ellipticity parameter. Abscissa is dimensionless \& the ordinate is in GeV$^{-2}$ units. }
\end{figure}
\medskip
\par\noindent Now, we turn our attention back to the ellipticity parameter given in equation no. (\ref{eq:ellp}). Being a small angle, we have relegated the tangent as equivalent to its angular argument. The regression analysis thus to be undertaken is between circular and linear random variables, lying on LHS (ellipticity parameter) and RHS (magnetic field) respectively. Suffice it to say that the LHS is readily read off from the table no. (\ref{bigtable}). The result of correlation study is given in the table no. (\ref{regrem})

\begin{table}
\begin{center}
\begin{tabular}[h]{lccccr}
Coefficients & Mean & Std.-Error & F-statistics & t-value & Pr$(>|t|)$ \\
 \hline
 Slope & 7.471e-38 & 2.251e-38 & 11.0164 & 3.319 & 0.000964\\
 \hline
 \end{tabular}
 \caption{Regression Result for the mass of the pseudoscalar}
 \label{regrem}
\end{center}
\end{table}

\subsection{Case - II: Vacuum birefringence
Only}
Here if we assume the mixing to be absent then we get $\theta = 0$ and hence we get the circular polarisation as
\begin{equation}
\rm{V}\!\!=\!\! 2\left(\rm{sin}\left[\left(k_{\perp} - k_{+}\right) z\right] \right) <A^{*}_{||}(0)A_{\perp}(0)>
\label{circt}
\end{equation}
Here, we need to evaluate only one argument and it is the same as given above:
\begin{equation}
k_{\perp} - k_{+} = {1\over 2\omega}\;\left\lbrace\,-3\xi\,\right\rbrace
\label{simt}
\end{equation}
We note, that, the mass of the pseudoscalar cancels and the mixing term is assumed zero. We see, that, the circular polarisation has now become inversely proportional to frequency assuming the argument to be considerably small as in the other case. Given that no circular polarisation would be produced in this case, it would be uninteresting to ponder over here.
\subsection{Case III: Limiting Case}
\par\noindent
We note the essential non-linearity resulting from the two effect taken in conjunction. Since we can not just add the two effects separately, even if they both are small and perturbative, to obtain the final result. We also note that the two effects shall be competing with each other when the following condition is met
$$\omega\xi \approx g\mathcal{B}$$
Leaving aside the numerical prefactors -tentatively we see that, unless the  value of magnetic field is much larger than normal pulsars (as in magnetars) and the beam frequency used in experiment is quite high (unlike the present case), even the modest values coming from astronomical bounds on pseudoscalar may not be comparable with the vacuum birefringence effect \& the former is in fact larger in effect.
\subsection{Case IV: General Case\cite{mig17}}
\par\noindent
Here in this subsection we calculate the amount of circular polarisation, vide the stokes parameter $V$, without resorting to any of the approximations made in the preceding two subsections, for completeness. Here the expression for the $V$ becomes
\begin{equation}
\rm{V}\!\!=\!\! \left\{\rm{cos}^2\left[\frac{g\mathfrak{B}\omega}{7 \xi\omega^2\sin^2\alpha + \omega_p^2 +  m_\mathfrak{a}^2}\right]\rm{sin}\left[\left(k_{\perp} - k_{+}\right) z\right] + \rm{sin}^2\left[\frac{g\mathfrak{B}\omega}{7 \xi\omega^2\sin^2\alpha + \omega_p^2 +  m_\mathfrak{a}^2}\right]\rm{sin}\left[\left(k_{\perp} - k_{+}^{'}\right) z\right]\right\} <A^{*}_{||}(0)A_{\perp}(0)>
\end{equation}
Hence at lowest order the first term, in $(\theta \ll 1)$ limit, would not change anything from what the first term, for the case of the pure mixing effect, did. But the second term, even at the lowest order shall render the expression for $V$ qualitatively different from what is was then at pure mixing effect. Needless to say that pure vacuum birefringence effect does not match, even qualitatively, with any of them, either.
For completeness, we write down the values of the wave vectors again.
\begin{equation}
k_{\perp} - k_{+} = {1\over 2\omega} \left\lbrace\,\xi\left(4\cos^2\alpha-7\sin^2\alpha\right)+\left[\frac{\left(g\mathfrak{B}\omega\right)^2}{7 \xi\omega^2\sin^2\alpha +  m_\mathfrak{a}^2}\right]\,\right\rbrace\;
k_{\perp} - k_{+}^{'} = {1 \over 2\omega} \left\lbrace\,4\xi\cos^2\alpha+m_\mathfrak{a}^2-\left[\frac{\left(g\mathfrak{B}\omega\right)^2}{7 \xi\omega^2\sin^2\alpha +  m_\mathfrak{a}^2}\right]\,\right\rbrace 
\label{simx}
\end{equation}
Thus far we have only shown the difference of results of all three separate cases in terms of the $V$ parameters depicting circular polarisation. This can be done with other two linear polarisation degrees of freedom, too. We leave this for a future endeavour.

\section{Result}
\label{res}
By the careful analysis of \cite[containing 600 Pulsar polarisation data, of which 537 are used here]{parkesix18} and that of \cite[containing 47 absolute PPAs for pulsars, 30 of which are common to the above]{ranki15}, we came to the following result given in table no. (\ref{rsl}).
\begin{table}[!h]
   \centering 
 {\rowcolors{3}{green!80!yellow!50}{green!70!yellow!40}
\begin{tabular}{ |p{3cm}|p{4cm}|p{3cm}|  }
\hline
\multicolumn{3}{|c|}{Results Obtained} \\
\hline
Parameters & Values  & Significance Level \\
\hline
$g_{\phi\gamma\gamma}$ & $4.903\times10^{-13}\;{\rm GeV^{-1}}$ & $ \simeq 0.001\%$ \\
$m_{\mathfrak{a}}$ &  $2.733\times10^{-10}\;{\rm eV}$  & $ < 0.1\%$ \\
\hline
\end{tabular}
}
\caption{The result of this analysis}
\label{rsl}
\end{table}
We however note that more data samples on absolute PPAs are required to obtain a more statistically significant result on the coupling, which is deduced, from this parameter. Currently a little over fifty (50) pulsars are amenable to this type of absolute PPA studies. The second quantity, namely the mass, has the numbers ($>$ 500) on its side. Nonetheless, its extraction from the ellipticity parameter, in turn, hinges on the coupling value, indirectly affecting the confidence interval found from the population. Also, for the sake of thoroughness we mention that the degree of linear polarisation is claimed to be dependent on frequencies in which they are observed \cite{han15}. The PPAs that are quoted in \cite[Absolute PPAs for 47 pulsars]{ranki15}, are for various radio frequencies, e.g. 327 MHz, 691 MHz,  3.1 GHz etc., including that of 1.4 GHz, which corresponds to 21 cm. Since, there has been no connection to PPAs are made with frequency, to our knowledge to this date, we did not investigate this further. 

\section{Discussion \& Outlook}
Taking advantage of new age of data explosion arising out from newer observational techniques and that of machine tools, we tried to estimate pseudoscalar particle mass and its coupling to photons. The results thus obtained do not match any standard axion models such as DFSZ or KSVZ etc. Hence these finding must be accommodated in the fold of axion like particles (ALPs) outside of the QCD realm. Surprisingly, our bottom up study, has automatically, led us to values, that are comparable and between the contemporary theories on cosmic axion background radiation (CAB), leading to soft X-ray excesses observed from comma cluster \cite{conlo13} \& that of the extra-galactic background light (EBL) to ALPs conversion \& oscillation, leading to an observed anomalous $\gamma$ ray transparency of the universe \cite{horn13}. Fortunately, the previous constraints  set on the mass and coupling of pseudoscalars, either by the changes in of quasar polarisation, hypothetically by ALPs \cite{payezcap12}, or by the $\gamma$ ray burst SN1987A \cite{payezcap15}, occurring through a so called ALPs burst,  are not in conflict with our results, either.
\par\noindent
As mentioned in section  no. (\ref{epppa}) a future  incorporation of vacuum birefringence effect into this study, may be performed, so as to see how the result on these estimates may change, for better or worse. These parameters may also be harnessed for devising CDM/WDM models and to obtain their relic densities.

\section*{Acknowledgement}
The authors acknowledge the software R \cite{rteam} for easing out the statistical analysis. SM is thankful of the help of K. Bhattyacharya regarding the same.

\bibliography{test,emeq}

\begin{thebibliography}{10}

\bibitem{Maiani86}
L.~Maiani; S. Petrozenio \&~E. Zavattini.
\newblock {\em Phys. Lett. B}, 175:359, 1986.

\bibitem{pec77}
R.~D. Peccei and H.~Quinn.
\newblock {\em Phys. Rev. Lett.}, 38:1440, 1977.

\bibitem{wilc78}
F.~Wilczek.
\newblock {\em Phys. Rev. Lett.}, 40:279, 1978.

\bibitem{wein78}
S.~Weinberg.
\newblock {\em Phys. Rev. Lett.}, 40:223, 1978.

\bibitem{kim79}
J.~E. Kim.
\newblock {\em Phys. Rev. Lett.}, 43:103, 1979.

\bibitem{sik83}
L.~Abbott and P.~Sikivie.
\newblock {\em Phys. Lett. B}, 120:133, 1983.

\bibitem{Das05}
S.~Das et. al.
\newblock {\em Jour. Cosmol. Astropart Phys.}, 06(002), 2005.

\bibitem{saral04}
Jain P.; Narain G. \&~Sarala S.
\newblock {\em MNRAS}, 347:394, 2004.

\bibitem{agarwal11}
Nishant Agarwal; Archana Kamal \&~Pankaj Jain.
\newblock {\em Phys. Rev. D}, 83:065014, 2011.

\bibitem{agarwal12}
Nishant Agarwal; Pavan K. Aluri; Pankaj Jain; Udit Khanna \&~Prabhakar Tiwari.
\newblock {\em The European Physical Journal C}, 72:1928, 2012.

\bibitem{jain13}
Prabhakar Tiwari \&~Pankaj Jain.
\newblock {\em International Journal of Modern Physics D}, 22:50089, 2013.

\bibitem{tiwari12}
Prabhakar Tiwari.
\newblock {\em Phys. Rev. D}, 86:115025, 2012.

\bibitem{tiwari16}
Pankaj Jain \&~Prabhakar Tiwari.
\newblock {\em Mon. Not. R. Astron. Soc.}, 460(3):2698–2705, 2016.

\bibitem{Das08}
S.~Das et. al.
\newblock {\em Pramana}, 70:439, 2008.

\bibitem{payezprd12}
A.~Payez.
\newblock {\em Phys. Rev}, 85:087701, 2012.

\bibitem{payez11}
A.~Payez; J.R. Cudell \&~D. Hutsemékers.
\newblock {\em Phys.Rev. D}, 84:085029, 2011.

\bibitem{vogel17}
G.~Raffelt H.~Vogel {A. Kartavtsev}.
\newblock {\em JCAP01(2017)024}, 01:024, 2017.

\bibitem{pelgrims15}
Vincent Pelgrims \&~Damien Hutsemekers.
\newblock arxiv:1503.03482.

\bibitem{pelgrims16}
Vincent Pelgrims \&~Damien Hutsemekers.
\newblock arxiv:1604.03937.

\bibitem{payez13}
A.~Payez.
\newblock {Patras Workshop on Axions, WIMPs and WISPs}.
\newblock In {\em {arXiv:1309.6114}}, 2013.

\bibitem{payez14}
Alexandre Payez; Carmelo Evoli; Tobias Fischer; Maurizio Giannotti; Alessandro
  Mirizzi;~Andreas Ringwald.
\newblock {\em Jour. Cosmol. Astropart Phys.}, 1502(006), 2014.

\bibitem{lamy00}
D.~Hutsemékers \&~H. Lamy.
\newblock arXiv:astro-ph/0012182, December 2000.

\bibitem{hut98}
Hutsemekers D.
\newblock {\em Astronomy \& Astrophysics}, 332:410, 1998.

\bibitem{hut01}
Hutsemekers D.;~Lamy H.
\newblock {\em Astronomy \& Astrophysics}, 367:381, 2001.

\bibitem{hut05}
D.~Hutsemekers; R. Cabanac; H. Lamy;~D. Sluse.
\newblock {\em Astron.Astrophys.}, 441:915, 2005.

\bibitem{hut14}
Damien Hutsemekers; Lorraine Braibant;~Vincent Pelgrims and Dominique Sluse.
\newblock {\em Astron. Astrophys.}, 572(A18), 2014.

\bibitem{hut10}
D.~Hutsemekers; B. Borguet; D. Sluse;~R. Cabanac and H.~Lamy.
\newblock {\em Astron. Astrophys.}, 520(L7), 2010.

\bibitem{jackson07}
N.~Jackson; R. A. Battye; I. W. A. Browne; S. Joshi; T. W.~B. Muxlow and P.~N.
  Wilkinson.
\newblock {\em Mon. Not. R. Astron. Soc.}, 376:371–377, 2007.

\bibitem{jagan16}
A.~R.; Jagannathan~P. Taylor.
\newblock {\em Mon. Not. R. Astron. Soc.}, 459(1):L36–L40, 2016.

\bibitem{csaki03}
C.~Csaki et. al.
\newblock {\em Jour. Cosmol. Astropart Phys.}, 0305(005), 2003.

\bibitem{cudell08}
D.~Hutsemékers; J.R. Cudell \&~A. Payez.
\newblock {Invisible Universe International Conference}.
\newblock In {\em {AIP Conf. Proc.}}, volume 1038, page 211, 2008.
\newblock arXiv:0805.3946.

\bibitem{quinn77}
R.~D. Peccei \&~H. Quinn.
\newblock {\em Phys. Rev. D}, 16:1791, 1977.

\bibitem{pich95}
A.~Pich.
\newblock arXiv:hep-ph/9505231v1, May 1995.

\bibitem{dine00}
M.~Dine.
\newblock arXiv:hep-ph/0011376v2, November 2000.

\bibitem{sikivie85}
P.~Sikivie.
\newblock {\em Phys. Rev. D}, 32:2988, 1985.

\bibitem{turner94}
E.~Kolb \& M.~S. Turner.
\newblock {\em {The Early Universe}}.
\newblock Westview Press, 2nd ed. edition, 1994.
\newblock Chap 10.

\bibitem{srednicki81}
W.~Fischler \& M.~Srednicki {M. Dine}.
\newblock {\em Phys. Lett. B}, 104:199, 1981.

\bibitem{zakharov80}
M.~A. Shifman; A. I. Vainshtein \& V.~I. Zakharov.
\newblock {\em Nucl. Phys. B}, 166:493, 1980.

\bibitem{sengupta99}
P.~Majumdar \&~S. Sengupta.
\newblock {\em Class. Quant. Grav.}, 16:L89, 1999.

\bibitem{sen01}
A.~Sen.
\newblock {\em Int. Jour. Mod. Phys. A}, 16:4011, 2001.

\bibitem{bell_68}
A.~Hewish; S. J. Bell; J. D. H. Pilkington; P. F. Scott; \& R.~A. Collins.
\newblock {Observation of a Rapidly Pulsating Radio Source}.
\newblock {\em Nature}, 217:709, 1968.

\bibitem{longair_94}
M.~S. Longair.
\newblock {\em {High energy astrophysics}}, volume~2.
\newblock Cambridge University Press, 1994.
\newblock p.99.

\bibitem{longair_96}
M.~S. Longair.
\newblock {\em {Our evolving universe}}.
\newblock CUP Archive, 1996.
\newblock p.72.

\bibitem{pulsar}
{Pulsar Properties}.
\newblock Webpage -
  https://www.cv.nrao.edu/course/astr534/PDFnewfiles/Pulsars.pdf.

\bibitem{lohfink_08}
Anne Lohfink.
\newblock {Pulsars}.
\newblock webpage - https://www.astro.umd.edu/~alohfink/seminar.pdf, November
  2008.

\bibitem{pul_hbk_05}
Duncan~Ross Lorimar and Michael Kramer.
\newblock {\em {Handbook of Pulsar Astronomy}}.
\newblock Cambridge University Press, 2005.

\bibitem{20_mil_pol_11}
W.~M.~Yan et. al.
\newblock {Polarization observations of 20 millisecond pulsars}.
\newblock {\em Mon. Not. R. Astron. Soc.}, 414:2087–2100, 2011.

\bibitem{multi_pol_pul_17}
Alice~K. Harding and Constantinos Kalapotharakos.
\newblock {Multiwavelength Polarization of Rotation-Powered Pulsars}.
\newblock 2017.
\newblock arXiv:1704.06183.

\bibitem{pul_mag_I_17}
{\em {New Advances in Pulsar Magnetosphere Modelling}}, {PoS}. SISSA, 2017.
\newblock arXiv:1702.00732.

\bibitem{pul_mag_II_16}
Vasily Beskin.
\newblock {Pulsar Magnetospheres and Pulsar Winds}, 2016.
\newblock arXiv:1610.03365.

\bibitem{parkesix18}
Simon Johnston and Matthew Kerr.
\newblock {Polarimetry of 600 pulsars from observations at 1.4 GHz with the
  Parkes radio telescope}.
\newblock {\em Mon. Not. R. Astron. Soc.}, 474(4):4629–4636, March 2018.

\bibitem{wang12}
P.~F. Wang, C.~Wang, and J.~L. Han.
\newblock {Curvature radiation in rotating pulsar magnetosphere}.
\newblock {\em Mon. Not. R. Astron. Soc.}, 423(1):2464–2475, 2012.

\bibitem{ganga10}
R.~T. Gangadhara.
\newblock {\em The Astrophysical Journal}, 710:29–44, 2010.

\bibitem{ardavan08}
Houshang Ardavan, Arzhang Ardavan, Joseph Fasel, John Middleditch, Mario Perez,
  Andrea Schmidt, and John Singleton.
\newblock {A new mechanism for generating broadband pulsar-like polarization}.
\newblock Number~78, page~16. Sissa, 2008.

\bibitem{mckinn02}
Mark~M. McKinnon.
\newblock Statistical modeling of the circular polarization in pulsar radio
  emission and detection statistics of radio polarimetry.
\newblock 568:302--311, 2002.

\bibitem{diagtool13}
Phrudth Jaroenjittichai.
\newblock {\em {Pulsar Polarization As A Diagnostic Tool}}.
\newblock PhD thesis, Physics \& Astronomy(UoM), 2013.

\bibitem{stodol88}
Georg Raffelt and Leo Stodolsky.
\newblock {Mixing of the photon with low-mass particles}.
\newblock {\em Phys. Rev. D}, 37(5):1237, Mar 1988.

\bibitem{mandal09}
Avijit K. Ganguly; Pankaj Jain \&~Subhayan Mandal.
\newblock {\em Phys.Rev.D}, 79:115014, 2009.

\bibitem{melrose17}
D.B. Melrose and M.~Z. Rafat.
\newblock {Pulsar radio emission mechanism: Why no consensus?}
\newblock {\em J. Phys.: Conf. Ser.}, 932:012011, 2017.
\newblock doi :10.1088/1742-6596/932/1/012011.

\bibitem{ganguly12}
Avijit~K. Ganguly.
\newblock {Introduction to Axion Photon Interaction in Particle Physics and
  Photon Dispersion in Magnetized Media}.
\newblock In Eugene Kennedy, editor, {\em {Particle Physics}}, chapter~3, page
  49–74. InTech, April 2012.

\bibitem{cameron99}
R.~Cameron et~al.
\newblock {Search for nearly massless, weakly coupled particles by optical
  systems}.
\newblock {\em Phys. Rev. D}, 47(9):3707–3725, 1993.

\bibitem{gedal02}
M.~Gedalin, E.~Gruman, and D.~B. Melrose.
\newblock {\em Phys. Rev. Lett.}, 88:121101, 2002.

\bibitem{rank15}
Megan~M. Force, Paul Demorest, and Joanna~M. Rankin.
\newblock {Absolute Polarization Determinations of 33 Pulsars Using the Green
  Bank Telescope}.
\newblock {\em Monthly Notices of the Royal Astronomical Society},
  (4):4485–4499.

\bibitem{ranki15}
Joanna~M. Rankin.
\newblock {\em The Astrophysical Journal}, 804:112, 2015.

\bibitem{mig17}
R.~P. et.~al. Mignani.
\newblock {\em Mon. Not. R. Astron. Soc.}, 465:492, 2017.

\bibitem{han15}
P.~F. Wang, C.~Wang, and J.~L. Han.
\newblock {On the frequency dependence of pulsar linear polarization}.
\newblock {\em Mon. Not. R. Astron. Society}, 448(1):771--780, 2015.

\bibitem{conlo13}
J.~P. Conlon and M.~C.~D. Marsh.
\newblock {\em Phys. Rev. Lett.}, 111(15):151301, 2013.

\bibitem{horn13}
M.~Meyer, D.~Horns, and M.~Raue.
\newblock {\em Phys. Rev. D}, 87:035027, 2013.

\bibitem{payezcap12}
A.~Payez; J.R. Cudell \&~D. Hutsemékers.
\newblock {\em Jour. Cosmol. Astropart Phys.}, 1207(041), 2012.

\bibitem{payezcap15}
A.~Payez, C.~Evoli, T.~Fischer, M.~Giannotti, and A.~Mirizzi.
\newblock {Revisiting the SN1987A gamma-ray limit on ultralight axion-like
  particles}.
\newblock {\em Jour. Cosmol. Astropart. Phys.}, 006(02), 2015.

\bibitem{rteam}
{R Core Team}.
\newblock {\em R: A Language and Environment for Statistical Computing}.
\newblock R Foundation for Statistical Computing, Vienna, Austria, 2013.

\end{thebibliography}


\end{document}